\documentclass[conference]{IEEEtran}

\makeatletter
\def\ps@headings{%
\def\@oddhead{\mbox{}\scriptsize\rightmark \hfil \thepage}%
\def\@evenhead{\scriptsize\thepage \hfil \leftmark\mbox{}}%
\def\@oddfoot{}%
\def\@evenfoot{}}
\makeatother

\usepackage{cite}

\ifCLASSINFOpdf
\else
\fi

\usepackage{url}
\usepackage{amsmath}
\usepackage{algorithmic}
\usepackage{algorithm}
\usepackage{epstopdf}
\usepackage{epsfig}
\usepackage{graphics}
\usepackage{subfigure}
\usepackage{amsfonts}

\newcommand{\bs}[1]{\ensuremath\boldsymbol{#1}}

\begin{document}

\title{Locating Disruptions on Internet Paths through End-to-End Measurements}

\author{Atef Abdelkefi$^{1}$, Yaser Efthekhari$^{2}$, Yuming Jiang$^{1}$\\
%, Yuming Jiang$^{1}$, Alexandru Calu$^{1}$, Iaonnis Lambadaris$^{2}$, Amir Banihashemi$^{2}$\\
\IEEEauthorblockA{$^{1}$~Centre for Quantifiable Quality of Service (Q2S), Norwegian University of Science and Technology, Norway\\
$^{2}$~Department of Systems and Computer Engineering, Carleton university, Canada}\\
}

\maketitle

\begin{abstract}
In backbone networks carrying heavy traffic loads, unwanted and unusual end-to-end delay changes can happen, though possibly rarely. In order to understand and manage the network to potentially avoid such abrupt changes, it is crucial and challenging to locate where in the network lies the cause of such delays so that some corresponding actions may be taken. To tackle this challenge, the present paper proposes a simple and novel approach. The proposed approach relies only on end-to-end measurements, unlike literature approaches that often require a distributed and possibly complicated monitoring / measurement infrastructure. Here, the key idea of the proposed approach is to make use of \emph{compressed sensing} theory to estimate delays on each hop between the two nodes where end-to-end delay measurement is conducted, and infer critical hops that contribute to the abrupt delay increases. To demonstrate its effectiveness, the proposed approach is applied to a real network. The results are encouraging, showing that the proposed approach is able to locate the hops that have the most significant impact on or contribute the most to abrupt increases on the end-to-end delay.
%In a (e.g. backbone) network, unwanted abrupt end-to-end delay changes can happen, though possibly rarely. In order to understand and manage the network to potentially avoid such abrupt changes, it is crucial to locate where in the network they have happened so that some corresponding actions may be taken. To tackle this challenge, the present paper proposes a simple and novel approach. The proposed approach simply only relies on end-to-end measurements, unlike literature approaches that often require distributed and possibly complicated monitoring / measurement infrastructure. A key and novel idea of the proposed approach is to make use of compressed sensing to estimate delays on each hop between the two nodes where end-to-end delay measurement is conducted. To demonstrate its effectiveness, the approach is applied to a real network. The results are encouraging, showing that the proposed approach is  able to  locate the hops that have most significant impact on or contribute most to abrupt changes on the end-to-end delay.
\end{abstract}
\section{Introduction}
The networking service offered by an Internet Service Provider (ISP) should maintain the highest possible reliability, stability and performance. Unwanted abrupt end-to-end delay increases, called {\em disruptions} in this paper, are not appreciated by the users.\footnote{Disruptions might be mainly caused by congestion in intermediate nodes.} It is, therefore, important for ISPs to monitor the behavior of the Internet paths that are responsible for these disruptions, and identify the critical nodes on such paths potentially causing them.

In the literature, several approaches were proposed in order to identify disruptions and the critical nodes involved. For instance, the authors in \cite{Andersen01,Feamester03} collect active probes cooperatively from a set of well monitored nodes in the Internet to detect failures. They further centrally analyze sets of traceroutes, network topologies and routing data in a dedicated server to infer the location of failures in the network. In \cite{PlanetSeer04}, the authors combined active probes and passive measurements from a set of dedicated nodes to detect and pinpoint routing loops, path changes and path outages in the Internet. Although these techniques are able to locate failures in the network, they all require a sophisticated measurement infrastructure that is hardly available in practice.

In this paper we propose a simple approach for the identification of critical nodes in a network. The proposed approach relies only on end-to-end measurements that can be easily implemented. Indeed, we answer the following question: given a set of observed disruptions on an Internet path, how can we pinpoint responsible nodes using a simple and easy to implement end-to-end measurement infrastructure?

To tackle this question, we investigate the applicability of \emph{compressed sensing} (CS) theory \cite{CSD,CSC}. To the best of authors' knowledge, there is no previous work that applies CS theory for the purpose of locating critical nodes on a path with disruptions through end-to-end measurements.\footnote{Authors in \cite{Firooz10,Firooz11} applied CS in the context of network tomography to infer {\em link-level} parameters, such as the number of links in a path, through end-to-end probing. Their work is not able to find critical nodes in disrupted path, and hence, not applicable to our problem.}

CS is a rapidly growing area of research which exploits the fact that many large data-sets are comprised of only a few significant elements; the notion of \emph{sparsity}. In CS, a vector $X \in \mathbb{R}^n$, with only $k, k \ll n$, significant elements, is \emph{recovered} from a set of $m, k < m \ll n,$ linear and noisy measurements $Y$ ($Y = \bs{R} X + N$). The matrix $\bs{R}$, characterizes the linear transform and is referred to as the \emph{sensing matrix}, while the vector $N$ denotes the noise present in the measurements.

In our proposed approach, end-to-end probing is conducted, with which end-to-end delays are measured and recorded. The end-to-end delays serve as the measurement vector $Y$. Then, we assume that a mechanism exists to detect abrupt and unusual delay increases, henceforth referred to as ``anomalous delays'', in the recorded measured delay sequence. The mechanism may vary from a simple thresholding technique to more advanced \emph{Principal Component Analysis} (PCA) techniques \cite{Abdelkefi11}. The details of such techniques are out of the scope of this paper and are omitted due to space limitation. When the anomalous delays are known, the next step is to acquire information about the routers involved in a certain end-to-end delay measurement through end-to-end traceroute probing. This information enables us to construct a routing matrix that will serve as the sensing matrix $\bs{R}$ later on. With the knowledge of the measurement vector $Y$ and the sensing matrix $\bs{R}$, we focus our attention on finding the minimum number of nodes that might have contributed to the detected anomalous delays with high probability. The result is a low-complexity counting algorithm designed based on CS methodology. The proposed algorithm has been validated with simulation and data collected from a real backbone network and shows sufficiently good accuracy.

The paper is organized as follows. In Section~\ref{sec2} we introduce the measurement setup and dataset. In Section~\ref{sec3}, we present compressed sensing basics and the problem formulation. In Section~\ref{sec4} we introduce the CS-based algorithm. The validation is presented in Section~\ref{sec5} and we conclude in Section~\ref{sec6}.

\section{End-to-end Delay Measurement and Traceroutes} \label{sec2}

Throughout this paper, we use delay measurement data collected from a backbone network. The setup is shown in Fig. 1, which includes two measurement computers, called measurement end-systems, and the networks connecting them. The NTE network is a commercial network in Norway while AUCKLA is an education and research network in Auckland. These two networks are connected through the Internet backbone. In order to study the behavior of the measured Internet path, the measurement end systems are connected directly to the border gateway routers at both sides and are synchronized using the Network Time Protocol (NTP).% \cite{NTP}.

\begin{figure}[ht!]
%\begin{figure}
%\vspace{-2cm}
  \centering
    \includegraphics[width=0.4\textwidth]{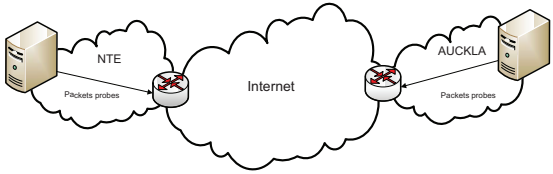}
%  \subfigure[CDF of the percentage of the energy captured for the measured aggregated delay]{\includegraphics[width=0.47\textwidth]{CDFenergyagdelay}}
%\vspace{-2cm}
  \caption{Measurement setup}\label{setup}
\end{figure}

Probe packets are sent between the two measurement end-systems in both directions every $10ms$. Every packet has a sequence number and is timestamped in both the transmitter and receiver, so that the one-way network delay, elements of the measurement vector Y, can be calculated. To generate, transmit and receive a continuous stream of probe packets, \emph{Real-time UDP Data Emitter} (RUDE) and \emph{Collector for RUDE} (CRUDE) tools are used \cite{RUDE}. The per packet delay and loss measurements were collected over one week in June 2010. % \cite{rudecrude}

In addition to the packet probes, traceroute probes were also collected to analyze the topology of the pairwise paths between the two end systems and infer the end-to-end connectivity. The sensing matric $R$ can be constructed using these measurement which can be later used to pinpoint the detected disruptions within a path.

\begin{figure}[ht!]
  \centering
    \includegraphics[width=0.25\textwidth]{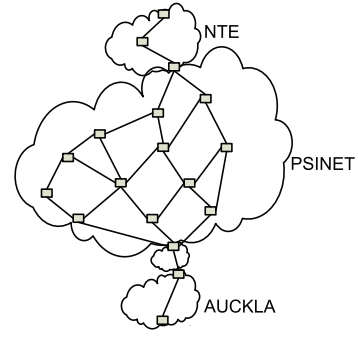}
%  \subfigure[CDF of the percentage of the energy captured for the measured aggregated delay]{\includegraphics[width=0.47\textwidth]{CDFenergyagdelay}}
  \caption{Measured commercial Internet path}\label{path}
\end{figure}

Fig. \ref{path} illustrates a typical topology for the commercial Internet path under consideration. Due to the load balancing mechanisms adopted in the network, multiple routes exist between the two end systems. We borrow Paxon's terminology\cite{paxon} and refer to the connectivity between the two end systems as \emph{virtual path} or simply a \emph{path} in the rest of the paper.

%Figure \ref{path} illustrates a typical topology for the commercial Internet path under consideration. Due to the load balancing mechanisms, multiple routes may exist between two end hosts. We borrow Paxon's terminology\cite{paxon} and refer to the connectivity between the end systems as virtual path or simply a path.

\section{Problem Formulation} \label{sec3}

Given the end-to-end delay measurements $Y$ and the sensing matrix $\bs{R}$, we will be using ideas from the theory of Compressive Sensing \cite{CSC,CSD} in order to detect the nodes responsible for the large delays observed in the data-set.

%\subsection{Background: compressed sensing}
\subsection{Compressed Sensing Basics}

The fundamental idea and appealing property of CS is to recover an $n$-dimensional vector $X$ through much smaller number $m$ of (noisy) measurements, exploiting the fact that $X$ is sparse (i.e. has much less than $n$ significant elements).\footnote{In CS it is also possible to have a signal which has a sparse representation in a transform domain. However, due to the special formulation of ours, we assume $X$ to be sparse itself.} The sampling process is essentially a linear transform and can be expressed as follows:

\begin{equation}
Y = \bs{R} X + N,
\label{CSForm}
\end{equation}
where $Y \in \mathbb{R}^{m}$ and $N \in \mathbb{R}^{m},$ $m \ll n,$ denote the measurement and the noise vectors, respectively. Considering the dimensions, to recover $X$ from $Y$, one has to solve an under-determined system of equations. While this may sound impossible, if $X$ is sparse and $\bs{R}$ satisfies the Restricted Isometry Property (RIP)\footnote{The RIP essentially requires the columns of $\bs{R}$ to be approximately orthogonal.} \cite{CSC,CSD}, CS guarantees the recovery of $X$ through solving the following optimization \cite{CSC,CSD}:
\begin{equation}\label{eql1}
\min\ \|X\|_{1},\ subject\ to\  Y= \bs{R} X + N,
\end{equation}
where the norm-1 of $X$, $\|X\|_{1}=\sum_{i}^{n} |X_{i}|$.
\subsection{Problem Formulation in Compressed Sensing Terms}
Let $X_i(t)$ and $Y_j(t)$ denote the delay at node $i$ and the $j$th anomalous end-to-end delay measured at time $t$, respectively. It is important to note that the value of $Y_j(t)$ is, in essence, the addition of delays at participating nodes $X_{i_1}, X_{i_2}, \cdots$, at times $t-\delta_{i_1}, t-\delta_{i_2}, \cdots$; i.e., $Y_j(t) = X_{i_1}(t-\delta_{i_1}) + X_{i_2}(t-\delta_{i_2}) + \cdots$. In practice, however, it is reasonable to assume an \emph{approximately stationary} model for the nodes in the network, so that, for example, the value of $X_{i_1}(t-\delta_{i_1})$ can be approximated by $X_{i_1}(t)$ with negligible distortion.\footnote{We will account for the distortion in the noise vector $N$ in Eq.~\ref{eq1}.} With this assumption, we let $Y(t) \in \mathbb{R}^{m}$ and $X(t) \in \mathbb{R}^{n}$ be the measured end-to-end delay vector and the vector of node delays at time $t$, respectively.

In practice, delays are probed every 10 ms. This time resolution results in tremendous amount of delay measurements $Y$. In order to reduce the amount of data, without loosing valuable information, we aggregate (sum-up) the measured per-packet end-to-end delays over time interval of duration $\delta$, referred to as \emph{bins}. This is motivated by the recent observation that aggregation can effectively make the disruption diagnosis of Internet paths more scalable while accurately capturing the unwanted abrupt changes \cite{Abdelkefi11}. For the aggregation on the measurements $Y$ to be effective, we need the following assumptions: 1) the traceroute does not change over the period $\delta$, and 2) the delay at nodes ($X_i(t)$) should also be aggregated over the same period $\delta$.

Considering an observation period $T$, not all $T/\delta$ time bins contain increased delays or otherwise anomalous behavior. Hence, $Y$ and $\bs{R}$ are the vector of aggregate end-to-end delays and the sensing matrix (constructed based on traceroutes) during the bins showing increased delay, respectively. The sensing matrix $\bs{R} \in \mathbb{R}^{m \times n}$ is a binary matrix, with elements $R_{ij}$ such that:

\begin{center}
$ R_{ij} = \begin{cases}
        1 & \text{if node } j \in i^{th} \text{ route}, \\
        0 & \text{otherwise}.
       \end{cases}$
\end{center}

Further, let $X_i$ denote the \emph{average} (aggregate) delay for node $i$ taken over all anomalous time bins within the observation period $T$. Arranging such delay per nodes in a vector $X$ for the interval $T$, we have the following linear model for the relation between the end-to-end ($Y$) and per node ($X$) delays:

\begin{equation}\label{eq1}
Y= \bs{R} X + N,
\end{equation}
where $N$ denotes the noise in the measurements, due to the averaging over node delays $X$ as well as the assumption that $X$ is stationary.

Following the notation introduced before, let $m$ and $n$ denote the number of anomalous time bins and the total number of nodes in their corresponding traceroute, respectively. In practice, it is known that only few nodes $k, k \ll n$, (within the Internet path) contribute to all disruptions during the observation window $T$. Therefore, the vector $X$ has only a few significant elements (high aggregate delays), and hence is considered \emph{sparse}. Henceforth, the nodes corresponding to significant elements of $X$ are referred to as ``critical nodes''.

Given $Y$ and $\bs{R}$, our goal in this paper is to find the minimum number of critical nodes and their location responsible for all anomalous time bins during $T$, i.e, recover the $k$ critical nodes of the vector $X$. Since the vector $X$ is sparse while the number of anomalous time bin $m$ is lower than the total number of nodes $n$, Compressed Sensing, is an appropriate theory that has been proven to solve such an underdetermined problem \cite{CSD}.

We will discuss in the next section, how to make use of CS theory to recover the vector $X$.

\section{The CS Algorithm} \label{sec4}

Recovery algorithms in CS are categorized based on the properties of the (sensing) matrix $\bs{R}$. Based on our formulation, $\bs{R}$ is binary, sparse and irregular in the sense that the number of $1$s in each row (column as well) might be different. Also, the columns of $\bs{R}$ might be highly correlated. Under such characteristics, $\bs{R}$ does not satisfy the RIP, unless the number of disruptions $m$ is increased to be in the order of $\Omega(k^{2})$ \cite{Gilber10}. This is achieved by increasing the observation window $T$, which may induce inaccurate results due to the increase of the number of critical nodes $k$ to recover. Therefore, CS recovery algorithms based on $\ell_1$ minimization (Equation \ref{eql1})\cite{CSC,CSD} are not applicable here.

We explore a different set of CS algorithms denoted as Verification-based \cite{SBB206,ZP07}. Verification-based (VB) recovery algorithms, inspired by decoding techniques in the context of channel coding, are known to perform well for the class of binary sparse sensing matrices \cite{EHBLISIT11,EHBL11}. Each VB algorithm, identifies and recovers the value of the elements in the support set through iterations. Once recovered, such values are fixed throughout the algorithm and help in recovering other support elements in further iterations. The major drawback of VB algorithms is their sensitivity to the noise in the measurements. Authors in \cite{EHBL11} discuss a thresholding technique to circumvent this issue. The performance of VB algorithms equipped with the thresholding technique depends mainly on the power of the signal elements in $X$ versus that of the noise (also known as SNR). Based on our empirical observations, such SNR assumptions might be well violated in the scenario under consideration in the paper.

Since we are only interested in the recovery of the location of the critical nodes and not their corresponding delay, we propose a ``hybrid counting'' recovery algorithm, inspired by the VB algorithms in the context of CS and the iterative bit flipping decoding algorithm in the context of LDPC codes \cite{LPDC}. The algorithm can be categorized as a VB recovery algorithm and aims at selecting nodes that are potential critical nodes in iterations.

The counting technique is inspired from a simple observation/assumption: since the set of critical nodes is considered sparse, each critical node is present in many anomalous traceroutes. The core of the counting technique is to count the number of traceroutes that pass from each node; the number of $1$s in each column of the matrix $\bs{R}$. If the sequence of counts have a large variance, then there is a high chance that the critical nodes we are looking for are among the columns with the highest count.

The algorithm works in three main phases: separation, identification and fusion. In what follows, we shall describe these phases.

In the separation phase, we divide the set of end-to-end delay measurements into disjoint subsets based on the value of the measurements. The measurements in each subset will have values close to other members of the subset, and apart from the delays in other subsets. For the measure of ``closeness'' we shall use the thresholding algorithm proposed in \cite{EHBL11}.

In each subset resulted from the separation phase, we shall apply the following steps of the identification phase:

\begin{enumerate}
\item Define a counter C and initialize it to zero.
\item Count the number of $1$s in the columns of the matrix.
\item Locate and save the index of the column with the highest count and all rows with a $1$ in this column.
\item Omit all the rows found in the previous step and increment counter C by one.
\item Go to step (2) unless all rows are deleted.
\item Store the counter C and all the pivotal columns of step (3).
\end{enumerate}

In the case of a tie at a specific step, the same step should be carried out for all the potential candidates.

When this phase terminates, it will list all the potential sets of critical nodes. The sets in the list may have different number of nodes as potential critical nodes. Because the number of critical nodes is supposed to be small and since our CS formulation aims at selecting the minimum number of critical nodes, within the list, we keep the sets with the minimum number of nodes (column indices) and proceed to the fusion phase.

In the fusion phase, we take the intersection between the results of identification phases for different groups of measurement from the separation phase. If the intersection consists of only one set, it is reported as the set of critical nodes. If not, we choose among the sets, one in which the nodes are less ``correlated''. The correlation between the nodes is defined as the dot product between their respective columns in the routing matrix $\bs{R}$.

For clarity, here we give an example of the counting algorithm in action.

{\bf Example:} Assume that the routing matrix $\bs{R}$ and the measurement vector $Y$ are as Tables \ref{rout_ex} and \ref{meas_ex}, respectively. The last row in Table \ref{rout_ex} indicates the number of $1$'s in the columns of $\bs{R}$. We number the rows and columns of $\bs{R}$ from top to bottom and from left to right, respectively.

\begin{table}[hb]
\caption{An instance of routing matrix with $13$ nodes and $7$ measurements. The last row indicates the number of $1$'s in each column.}
\label{rout_ex}
%\left
$\bs{R}$=[
\begin{tabular}{ccccccccccccc}
0&1&0&0&1&1&1&0&0&1&0&0&0\\
1&0&1&0&1&1&0&0&0&0&0&1&0\\
1&1&0&0&0&0&1&0&0&0&1&0&0\\
0&1&1&0&0&1&0&0&0&1&0&0&1\\
0&1&0&0&0&0&1&1&1&1&1&0&0\\
1&0&0&1&1&1&1&0&0&0&0&1&0\\
0&0&0&1&0&0&1&0&1&1&1&0&0\\
\hline
3&4&2&2&3&4&5&1&2&4&3&2&1\\
\end{tabular}
%\right
]
\end{table}

\begin{table}[hb]
\caption{An instance of the measurement vector $Y$.}
\label{meas_ex}
%\left
$Y$=[
\begin{tabular}{lllllll}
1930&1929&1930&1933&1933&1932&1934\\
\end{tabular}
%\right
]$^T$
\end{table}

Since the measurements are considered close in this case, the separation phase is ignored. The maximum count of $5$ happens at the 7th column. We save this column considering it as pointing to a potential critical node, and note that in this column, rows 2 and 4 are zero and the rest are $1$. According to step (4) of the identification phase, we delete all the rows except for these two. So, the remaining matrix has only two rows and is shown in Table \ref{rout1_ex}.

\begin{table}%
\caption{Reduced routing matrix in the second round of the identification phase.}
\label{rout1_ex}
%\left
$\bs{R}$=[
\begin{tabular}{ccccccccccccc}
1&0&1&0&1&1&0&0&0&0&0&1&0\\
0&1&1&0&0&1&0&0&0&1&0&0&1\\
\end{tabular}
%\right
]
\end{table}

In the second round of the identification phase, the maximum count is $2$ and occurs at columns 3 and 6. Since the number of $1$'s in such columns equals the total number of rows, the algorithm shall stop at this stage and both columns 3 and 6 are reported as potential critical nodes for the second round.

Hence, the potential sets of critical nodes are: $\{7,3\},\{7,6\}$; first position (7) from the first round and the second position from the second round. All these sets report two nodes as the potential critical nodes ($k = 2$). In the fusion phase, we shall decide between the two sets. Note that the correlation of the 3rd and 7th columns is zero, while the correlation of the 6th and 7th columns is non-zero. Therefore nodes 3 and 7 are selected as critical nodes.

\section{Validation}\label{sec5}
In this section we will provide an evaluation of our proposed algorithm for critical nodes recovery. We conducted the evaluation of our critical node recovery algorithm based on both simulation and collection of dataset from real environment.
\subsection{Simulation}
In this section, we will provide numerical simulation results demonstrating the performance of our proposed algorithm. For that, we generate for a random binary routing matrix with $16$ rows (m=$16$) and $40$ columns (n=$40$), so that each row has exactly $6$, 1's ($6$ nodes contribute to each traceroute measurement).

As the number of critical nodes highly affects the recovery performance of any CS-based algorithm, the target of this simulation scenario is to study the detection and false positive rates of the proposed algorithm, while varying the number of the critical nodes. However, we additionally noticed that for a fixed sensing matrix $R$, the position of the critical nodes in the topology may dramatically change the detection and false positive rates. For example the recovery performance of the algorithm depends on the correlation between the columns of the routing matrix corresponding to the critical nodes. It also depends on the number of occurrences of a critical node in the traceroute measurements, i.e, the number of 1s in the columns of the routing matrix.

To capture the effect of all such parameters in the detection and false positive rate of the proposed algorithm, we fixed the routing matrix in the simulation, while varying the number and the position of the critical nodes among the set of all nodes. Then, for each fixed number of critical nodes, we found the average detection and false positive rates of the algorithm, while the average is taken over all possible positions of the critical nodes. As the number of critical nodes is assumed to be small compared to the total number of nodes, we choose $10$ as the maximum number of critical nodes in our simulation, corresponding to $25\%$ of the total set of nodes. Additionally in our simulations, we assigned a much higher delay to the critical nodes compared to the ordinary ones ,i.e, the delay at critical nodes is chosen to be three times higher.

\begin{figure}
  \center
  \subfigure[Detection rate as a function of the number of critical nodes]{\includegraphics[width=0.35\textwidth]{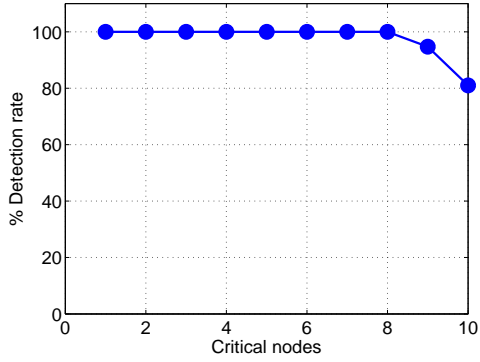}}
  \subfigure[False positive rate as a function of the number of critical nodes]{\includegraphics[width=0.35\textwidth]{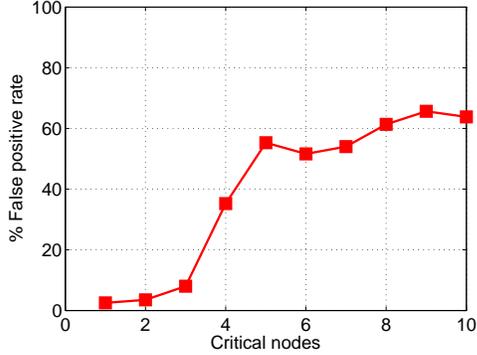}}
  \caption[width=0.23]{Algorithm performance}
  \label{algoperf}
\end{figure}

Figure \ref{algoperf} illustrates the detection and false positive rates as a function of the number of critical nodes. It shows that as the number of critical nodes increases, the detection rate decreases while the false positive rate increases. Figure \ref{algoperf} (a) shows that the algorithm maintains a $100\%$ detection rate if the number of critical nodes does not exceed 8 (out the $40$ nodes). Although relatively high, the detection performance degrades as the number of critical nodes increases until it reaches $81\%$ when the number of critical nodes is equal to $10$. The false positive rate on the other hand increases as the number of critical nodes increases. As can be seen, for small number of critical nodes (less than 4), the false positive rate is very low (does not exceed $5\%$). However, increasing the number of critical nodes further, increases the false positive rate dramatically to approximately $65\%$.

\subsection{Real environment}
In this section we discuss the validation framework which is based on a collected dataset from a real network as discussed in section \ref{sec2}. Unfortunately, validating the framework with a reasonable confidence in real environment is notoriously difficult due to many inaccuracies. For instance, traceroute data can be inaccurate since loops, cycles and diamonds may exist in the topology \cite{Augustin}. In addition, missing or false positive links might be reported due to the load-balancing nodes \cite{Augustin2}. Moreover, delays reported in traceroute measurement are Round-Trip Time delays (RTTs) which do not give a clear clue on the one-ways (forward/reverse path) where disruptions are experienced.

To fade the effect of such inaccuracies, we set up tarceroute-based probes with a high granularity ($\delta = 2$ min) together with the delay measurement infrastructure shown in Fig. \ref{setup}. In addition, we assume that an abrupt increase in the measured RTT is more likely due to a problem in the forward path \cite{Pucha07}, mainly due to the fact that most routes in commercial Internet tend to be asymmetric, thus disruptions in the forward path do not usually correlate with those of the reverse path.

\begin{figure}
  \centering
    \includegraphics[width=0.47\textwidth]{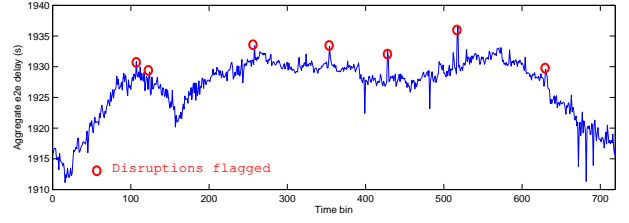}
%  \subfigure[CDF of the percentage of the energy captured for the measured aggregated delay]{\includegraphics[width=0.47\textwidth]{CDFenergyagdelay}}
  \caption{Aggregate end-to-end delay variation NTE-AUCKLA, $29/06/2010$}\label{e2evar}
\end{figure}

In our validation framework we locate critical nodes on the measured path NTE-AUCKLA during one of the measurement days. June 29th 2010 was chosen due to the relatively high number of disruptions detected. The delay time series is shown in Fig. \ref{e2evar}. For validation purposes, the end-to-end delays with variations exceeding the standard deviation $\sigma$ of the delay time series are chosen as anomalous delays. Since this selection criterion is not the main focus of this paper, further discussion about the chosen method is omitted.

% With this case, a time bin $i$ is flagged anomalous only if the aggregate delay variation at this time bin satisfies $\tau_{i}=d_{i}-d_{i-1} > 4.701 s$, where $4.701 s$ is the standard delay deviation for the chosen delay.

As shown in Fig. \ref{e2evar}, $7$ time bins were flagged as ``anomalous" during this one day observation window. The end-to-end (virtual) path consists of $24$ hops. Accordingly, the routing matrix $\bs{R}$ is a $7\times 24$ binary matrix and $Y \in \mathbb{R}^{7}$ is the vector of aggregate delay during the $7$ ``anomalous" time bins. The proposed counting algorithm reported two hops as the critical nodes. We denote their IP addresses by $IP_{1}$ and $IP_{2}$. $IP_1$ and $IP_2$ are indeed the nodes $3$ and $7$ of the example above, respectively. In the example, however, we reduced the size of the matrices for demonstration purposes.

\begin{figure}
  \centering
    \includegraphics[width=0.47\textwidth]{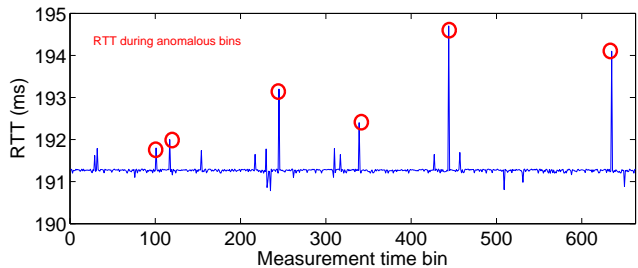}
%  \subfigure[CDF of the percentage of the energy captured for the measured aggregated delay]{\includegraphics[width=0.47\textwidth]{CDFenergyagdelay}}
  \caption{RTT variation for $IP_{2}$ during one day observation window}\label{RTTvar}
\end{figure}

Our validation methodology is mainly based on tracking the measured RTT variation during the observation window T. Particularly, the measured RTTs during the detected ``anomalous" time bins were compared to the average RTT within a one day observation window. Fig. \ref{RTTvar} illustrates an example of the validation methodology for $IP_{2}$. The figure shows that the average RTT between the source and $IP_{2}$ is around $191.275ms$ when the path is stable, i.e. no disruptions experienced. In addition, RTT suddenly increases in the order of \emph{few milliseconds} ($1$ to $4$ ms) during time bins when disruptions occur. The same behavior for $IP_{2}$ is observed in conjunction with the other $6$ time bins when disruptions are detected. This confirms that $IP_{2}$ was indeed a critical node.
However, it is worth noting that the measured RTT, may experience additional abrupt variations in time bins when the aggregate delay looks stable, i.e, non-flagged time bins. This is likely due to the inaccuracy in detecting disruptions in the delay time series, or simply due to a set of transient instabilities experienced by the node causing minor effects on the aggregate end-to-end delay.
%Nevertheless, fortunately, we are able to identify anomalous {\em patterns}, i.e, abrupt variations in the measured RTT, for the reported IP addresses (or routers) during the flagged anomalous time bins.

In what follows we validate the choice of $IP_{1}$ as a critical node. Table \ref{rttcomp} presents how the measured RTT varies between normal and flagged anomalous time bins, for $IP_{1}$ (as well as for $IP_{2}$). It shows an abrupt increase in the RTT between the source and $IP_{1}$ estimated in the order of $2\ ms$, during the flagged ``anomalous" time bin. Although not contributing to many of the disruptions, the high intensity of the disruptions induced by $IP_{1}$ is likely to be the reason why it is reported as critical.

%\begin{center}
\begin{table}
\caption{Comparison of avg RTTs} % within normal and anomalous time bins}
\label{rttcomp}
\begin{tabular}{|l|l|l|}
  \hline
% \multicolumn{6}{|c|}{traceroute to $IP_{Auckla}$, 30 hops max, 60 byte packets} \\
&Anomalous time bin&Non anomalous Time bin\\ \hline
$IP_{1}$ Average RTT&131.525 ms &129.079 ms\\ \hline
$IP_{2}$ Average RTT&193.033 ms&191.275 ms \\
 \hline
\end{tabular}
\end{table}
%\end{center}

%Further investigation has been done for the entire week of measurement period, where the same validation process was followed. Our framework shows appealing results with less than $15\%$ ($2$ out of $14$) of the identified critical IPs were reported false positive.

\section{Conclusion}\label{sec6}
This work presented a new and promising approach to infer ``critical" nodes responsible for a set of observed disruptions on an Internet (virtual) path, based on end-to-end measurements. While the end-to-end measurement technique has been widely adopted due to its simplicity, it offers a very limited overview of path characteristics. We show in this work that when combined with compressed sensing, end-to-end measurements can potentially be utilized effectively to infer the behavior of ``critical" nodes within the measured path. Particularly, we propose a simple counting algorithm inspired by compressed sensing and channel coding. The results based on real measurement data, were encouraging. Despite the challenges in traceroute analysis, mainly due to the measurement artifacts, the proposed algorithm was able to accurately infer the set of critical nodes for the entire measurement period. The low complexity of the proposed method as well as its relative accuracy, points to an interesting methodology for network disruptions diagnosis which we hope will motivate further research in this area.

%Future work will involve the detailed analysis of these questions.

\end{document}